\documentclass[aps,prl,preprint,superscriptaddress,showpacs,showkeys]{revtex4}
\usepackage{graphicx}
\usepackage{graphicx}
\usepackage{dcolumn}
\usepackage{color}
\usepackage{mathrsfs}
\usepackage[breaklinks,colorlinks = true,linkcolor = blue,
urlcolor=blue,citecolor=blue]{hyperref}
\usepackage{amsmath}
\usepackage{soul}

\begin{document}
\title{\bf Magnetic Phase Transitions in the NdFe$_3$(BO$_3$)$_4$ multiferroic}
\author{G.A.~Zvyagina}
\affiliation{B.I.~Verkin Institute for Low Temperature Physics and
Engineering of the National Academy of Sciences of Ukraine,
Kharkov, 61103, Ukraine}
\author{K.R.~Zhekov}
\affiliation{B.I.~Verkin Institute for Low Temperature Physics and
Engineering of the National Academy of Sciences of Ukraine,
Kharkov, 61103, Ukraine}
\author{I.V.~Bilych}
\affiliation{B.I.~Verkin Institute for Low Temperature Physics and
Engineering of the National Academy of Sciences of Ukraine,
Kharkov, 61103, Ukraine}
\author{A.A.~Zvyagin}
\affiliation{Max-Planck-Institut f\"ur Physik komplexer Systeme,
01187 Dresden, Germany}
\affiliation{B.I.~Verkin Institute for Low
Temperature Physics and Engineering of the National Academy of
Sciences of Ukraine, Kharkov, 61103, Ukraine}
\author{I.A.~Gudim}
\affiliation{L.V.~Kirensky Institute for Physics, Siberian Branch
of the Russian Academy of Sciences, 660036, Krasnoyarsk, Russian
Federation}
\author{V.L.~Temerov}
\affiliation{L.V.~Kirensky Institute for Physics, Siberian Branch
of the Russian Academy of Sciences, 660036, Krasnoyarsk, Russian
Federation}
\date{\today}

\begin{abstract}
Low temperature studies of the behavior of the sound velocity and attenuation
of acoustic modes have been performed on a single crystal NdFe$_3($BO$_3$)$_4$.
Transitions of the magnetic subsystem to the antiferromagnetically ordered
state at $T_N \approx 30.6$~K have been revealed in the temperature behavior 
of the elastic characteristics. The features in the temperature behavior of
elastic characteristics of the neodymium ferroborate and its behavior in the
external magnetic field, applied in the basic plane of the crystal, permit us
to suppose that the transition to an incommensurate spiral phase is realized
in the system. This phase transition behaves as the first order one. $H-T$ 
phase diagrams for the cases ${\bf H} \parallel {\bf a}$ and ${\bf H} 
\parallel {\bf b}$ have been constructed. The phenomenological theory, which 
explains observed features, has been developed.
\end{abstract}


\maketitle

\section{Introduction}

Rare-earth ferroborates with the common formula RFe$_3$(BO$_3$)$_4$ (R=Y,
La-Nd; Sm-Er) are studied intensively during last years, because they possess a
set of interesting optical, magnetic, and magneto-electric properties.
Belonging of these crystals to the class of multiferroics makes them very
perspective subjects for the creation of new multifunctional devices \cite{1}.

Magnetic properties of the rare-earth ferroborates are determined by the
presence there of two types of magnetic ions, which belong to $3d$ and $4f$
elements. The specifics of their magnetic properties is defined, first, by the
behavior of the magnetic iron subsystem, and second, by the characteristic
features of the electron structure of the rare-earth ion, which is formed by
the crystalline electric field, and, naturally, by the $f-d$ interaction. In
all compounds of this family the antiferromagnetic ordering of the iron
subsystem takes place at low temperatures (30~K $\le T_N \le 40$~K). The
question about the state of the rare-earth subsystem is still under debates.
In the majority of papers it is supposed that the rare-earth subsystem is
magnetized due to the $f-d$ interaction. The temperature of the
antiferromagnetic transition weakly depends on which rare-earth ion forms the
ferroborate. However, the ions R$^{3+}$ bring an essential contribution to the
orientation of the Fe$^{3+}$ magnetic ions in the ordered state, and the
magnetic anisotropy. Depending on the rare-earth ion, the compounds can be
antiferromagnets of ``easy-axis'' or ``easy-plane'' type, or can be
spontaneously transferred between those types . The concurrence between the
contributions of two magnetic systems in the formation of the magnetic
structure defines the existence of phase transitions, which take place with
the change of the temperature, as well as with the change of the external
magnetic field  \cite{1,2}.

The presence of the essential coupling between the magnetic, electric, and
elastic subsystems of ferroborates yields to the onset of multiferroelectric
effects, which reveal themselves most sharply in the vicinities of the
spontaneous and magnetic field-induced phase transformations. This is why,
the investigation of the elastic properties of such compounds in the
vicinities of phase transitions is very interesting.

The present study is related to the investigation of the magneto-elastic
effects in the ferroborate NdFe$_3$(BO$_3$)$_4$.

\section{Structure, magnetic, and magnetoelectric properties of the 
neodymium ferroborate}

The crystal structure of RFe$_3$(BO$_3$)$_4$ compounds belongs to the
structural type of the natural mineral huntite CaMg$_3$(CO$_3$)$_4$ that
crystallizes in the space group R32 of the trigonal system. In this structure,
chains of edge-sharing FeO$_6$ octahedra run along the trigonal ${\bf с}$-axis,
being interconnected by isolated RO$_6$ trigonal prisms and BO$_3$ triangles
\cite{3}. Within this family, the members with “heavy” rare-earth ions from
Eu to Yb, and also Y, experience a first order structural transformation to a
less symmetric P312$_1$ phase, which also belongs to the trigonal type
\cite{4}. According to the data \cite{5} NdFe$_3$(BO$_3$)$_4$ does not undergo
the symmetry reduction and remains in the R32 space group down to 2~K.

Magnetic properties of NdFe$_3$(BO$_3$)$_4$ have been explored in detail by the
presernt time. In particular, investigations of magnetic and thermal
properties of this compound have evidenced that below the temperature $T_N \sim
30$~K its' magnetic structure was antiferromagnetic with the magnetic moments
of iron oriented in the basic ${\bf ab}$-plane. Thereby, NdFe$_3$(BO$_3$)$_4$
has been presumed as a magnet with an ``easy-plane'' magnetization anisotropy
\cite{6,7}. Authors of those papers have believed that the subsystem of Nd
remained paramagnetic. The magnetic Nd moments magnetized by the effective
field developed by the ordered subsystem of iron ions also lay in the
${\bf ab}$-plane. The assumption about the existence of three kinds of domains
with vectors of antiferromagnetism oriented along the C$_2$ binary axes in zero
magnetic field has brought out. On one of the domains the direction of the
vector of antiferromagnetism coincides with the crystallographic
${\bf a}$-axis. Small jumps of the magnetization in the magnetic field applied
within the basic plane have been registered \cite{7}. These peculiarities (at
$H_{sf}$=1~T and $T=$2~K) have been related to the spin-flop transition. As
a result, both Fe and Nd moments have turned perpendicular to the direction of
the external field within the basic plane \cite{8}. The same conclusion has
been obtained in \cite{5}. The detailed analysis of the magnetically
diffracted intensities as a function of the magnetic field and azimuth angle
in the hard X-ray scattering experiments in NdFe$_3$(BO$_3$)$_4$ have
suggested that the magnetic moments of the Fe ions have aligned in the
crystallographic ${\bf a}$-axis, leading to a domain structure formation, since
there were three equivalent directions in the basic plane. On the other hand,
the last neutron-diffraction results \cite{9} have shown that the long-range
magnetic order observed in NdFe$_3$(BO$_3$)$_4$ below $T_N \approx 30$~K 
consisted of antiferromagnetic stacking along the ${\bf c}$ axis, where the 
magnetic moments of all three Fe$^{3+}$ sublattices and the Nd$^{3+}$ 
sublattice was aligned
ferromagnetically and parallel to the hexagonal basal plane. According to
\cite{9}, three magnetic Fe moments have equal magnitudes, and the angle
between the magnetic Fe and Nd moments was zero. Moreover, the phase
transition to the low-temperature incommensurate (IC) magnetic structure has
revealed at $T_{IC}=13.5$~K. The authors have supposed that in the
incommensurate magnetic phase the magnetic structure of NdFe$_3$(BO$_3$)$_4$
has transformed into a long-period antiferromagnetic helix \cite{9}.

Magnetic ordering has induced a variety of interesting features in this
material. Recently NdFe$_3$(BO$_3$)$_4$ have been reported to exhibit a
significant magnetodielectric coupling in an external magnetic field
\cite{10}. The stepwise onset of the large magnetically induced electric
polarization was found in the temperature range 4.2~K--25~K for fields $H_{cr}
\sim 1$~T parallel to the ${\bf a}$-axis. However, even for the absence of
the external magnetic field, in NdFe$_3$(BO$_3$)$_4$ the electric polarization
distinct from zero has been observed below the N\'eel temperature \cite{2}. A
clear relation between the magnetoelectric and lattice properties have been
shown in the magnetostriction measurements \cite{10}. Jump-like peculiarities
in the magnetostriction were also found in the same temperature and magnetic
field range, where peculiarities in electrical polarization have been
registered.

\section{Experimental setup}

Single crystals of NdFe$_3$(BO$_3$)$_4$ have been grown using a
Bi$_2$Mo$_3$O$_{12}$-based flux by the technique described in \cite{11}, and
have reached the sizes up to 10--12 mm. The seeds were obtained by the
spontaneous nucleation from the same flux. Single crystals were green in color
and had a good optical quality.

We worked with the crystal representing transparent hexagonal well faceted
prism in height about of 4 mm in a direction, close to an axis of symmetry of
the third order. Experimental samples with the characteristic sizes $\sim
1.9 \times 2.6\times  3.6$~mm (the sample 1) and
$0.8\times 1.2 \times 1.2$~mm (the sample 2) have been prepared. The samples
were oriented by means of the X-ray diffraction (the Laue method).

In the study the temperature and the magnetic field behavior of the velocities
and attenuations of the longitudinal and transverse acoustic modes in the
single crystal samples of NdFe$_3$(BO$_3$)$_4$ was performed. The range of the
temperature was 1.7~K--300~K, and the magnetic field was used up to 5.5~T.

The mesurements of the relative changes of the velocity and attenuation of
acoustic modes were performed using the automatized setup described in
\cite{12}. The working frequency was 54.3~MHz. The accuracy of the relative
measurements with the thickness of samples $\sim 0.5$~mm was about
$\sim 10^{-4}$  in the velocity and 0.05~dB in the attenuation.

\section{Temperature dependences of the sound velocity and
attenuation}

As is known, the neodymium ferroborate conserves the R32 symmetry at least up
to 2~K \cite{5}. Our experiments have shown that in the temperature range
300~K-30~K all longitudinal and transverse acoustic modes demonstrate the
standard monotonic behavior without any peculiarities. However, starting from
30.6~K till 10~K the velocity and attenuation of all studied modes behave in an
anomalous way.

The behavior of the transverse modes in this temperature range can be
characterized as follows. At $T_N \approx 30.6$~K the the jump-like decrease 
of the velocity is observed. For the following decrease of the temperature, the
velocity continues to decrease, reaching the minimum value in the vicinity of
$T \sim 25$~K. Up to $\sim 20$~K the value of the velocity is changed non
essentially. Then, one observes the essential growth of the velocity, so that
at 10~K it returns practically to its' value at $T_N \approx 30.6$~K, see Figs.
1a and 2a. We used the following notations in Figures: $\Delta s/s$ and
$\Delta \alpha$ are the relative changes of the velocity and attenuation of
sound waves with the wave vector ${\bf q}$ and the polarization ${\bf u}$,
which are spreading along the axes ${\bf a}$, ${\bf b}$ and ${\bf c}$ of the
standard for the trigonal crystal Decart system of co-ordinates (${\bf a}
\parallel$ C$_2$, ${\bf c} \parallel$ C$_3$).The dependences for different 
magnetic field values  are shifted for clarity along the ordinate
relative to one another in Figs.1-4.

In the case of the transvwerse C$_{44}$ (${\bf q} \parallel {\bf c}, {\bf u}
\parallel {\bf a}$) the growth of the velocity starts about $T \sim 14$~K, and
it is finished to $T \sim 7$~K  , Fig. 3a ($H=0$).

The behavior of the attenuation in this temperature region is seen in
Figs. 1b-3b ($H=0$). One can see that the peculiarities in the behavior
correlate with the anomalies observed in the sound velocity.

The positions of the peculiarities in the velocity and attenuation near
$T_N \approx 30.6$~K are not changed when the direction of the temperature
development. The behavior of the sound characteristics in the range 10~K
$\le T\le$ 25~K has the hysteresis character.

The longitudinal modes in the range 10~K $\le T\le$ 30.6~K behave in an
analogous way, though the scale of anomalies for them is an order of magnitude
weaker.

From 10~K and up to the lowest in our experiments temperature 1.7~K there are
no essential changes of the acoustic characteristics of the crystal.

We have found that in the discussed temperature range 10~K $\le T\le$ 30.6~K the
sample is in the inhomogeneous state (from the point of view of the sound
development). The mode C$_{44}$ ``feels'' this inhomogeneity most sharply.
The sound signal below $T_N \approx 30.6$~K becomes multiphase one, and
the temperature dependences become jagged. Clearly seen maxima appear in the
behavior of the attenuation, Fig. 4 the sample 1. Those maxima, most probably,
have the interference nature, since the change of the working frequency for
$\pm 1$~MHz causes to the redistribution  of their intensities and changes
their temperature position for 1-2~K, Fig. 4b.

We managed to obtain the better result (to exclude the influence of the
interference) by using of the significant change of sizes, and to some extend,
of the shape of the sample. In the sample 2 the sound signal was practically
single-phase one in all the range of temperatures, including the region of the
inhomogeneity  10~K $\le T\le$ 30~K. The dependences of the sound velocity and
sound attenuation appeared to be smooth, Fig. 4, the sample 2. The transverse
modes, which wave vector lies in the basic plane ${\bf ab}$ ``feel'' the
inhomogeneous state of a sample weaker. The temperature dependences of the
acoustic characteristics of the samples 1 and 2 are weakly different from
each other.

The magnetic field $H \le 1$~T, applied to the basic plane (${\bf H}
\parallel {\bf a}$, ${\bf H} \parallel {\bf b}$). practically does not change
the position of the feature at $T_N \approx 30.6$~K and of the low-temperature
wings of the temperature dependences of the velocity and attenuation. However,
the scale of anomalies becomes significantly smaller, so that in fields
$H > 1$~T only features at $T_N$ can be distinguished, Figs. 2,3.

Now, let us compare our results with the data for the observed specific heat,
magnetic susceptibility, X-ray and neutron scatterings. As it follows from
\cite{5,6,7,8,9}, NdFe$_3$(BO$_3$)$_4$ below the N\'eel temperature becomes
antiferromagnetically odered with the anisotropy of the ``easy-plane'' type
(the commensurate magnetic phase \cite{9}). The onset of a domain structure is
possible in that phase.

With the following decrease of the temperature, as it was shown in \cite{9,13},
the compound transfers to the incommensurate (IC) magnetic phase; as a result
the magnetic structure is transformed into a long-period antiferromagnetic
helix. The temperatures of such a phase transition, according to the neutron
scattering data, performed on the powder and single-crystal samples, are
determined as $T_{IC} \approx 19$~K and $T_{IC} \approx 13.5$~K, respectively.
At the same time, the authors of study used the nonresonant X-ray magnetic
scattering \cite{5} believe that the transition to the magnetic spiral
structure is not happened up to 2~K: Only antiferromagnetic domain structure
is formed.

It is known also \cite{2} that below $T_N \approx 30$~K the spontaneous
electric polarization appears in NdFe$_3$(BO$3$)$_4$. The temperature
dependences of the polarization along ${\bf a}$ and ${\bf b}$ axes manifest
peculiarities about 17~K, Fig. 4a.

Since there are no data about the presence of any structural transformations
in NdFe$_3$(BO$3$)$_4$, it is logical to relate the observed by us anomalies
in the behavior of the elastic characteristics in the range
10~K $\le T \le 30.6$~K with the transformation of the magnetic (and electric)
subsystem of the crystal. We believe that the anomalies at
$T_N \approx 30.6$~K are related to the transition to the magnetically ordered
state. The critical temperature of this transition, which is determined from
our experiments, correlates with the most frequently published value of the
N\'eel temperature for NdFe$_3$(BO$3$)$_4$.

Unfortunately, it is difficult, looking on the temperature dependences of the
sound velocity and attenuation, to answer the question about the realization
of the phase transition to the incommensurate magnetic phase in the system. In
the temperature dependeces below $T_N$ there is no well-localized peculiarity,
which can be defined as the temperature of the phase transition. However, for
all modes in the range 10~K $\le T \le 20$~K we registrate the anomalous
increase of the velocity, which tells us about the distortions of the crystal
lattice. Besides, the temperature dependences of the characteristics of the
C$_{44}$ mode, observed in the massive sample 1, have relatively sharp
($\sim 1.5$\%) step of the velocity, and associated with it maximum in the
attenuation namely at 13.5~K. In the thick sample 2 these anomalies become
more smooth, ``smeared out'' in temperature, Fig. 4. In our opinion, the most
probable reason for such a behavior, is the transition to the spiral magnetic
state.

One can present the following arguments in favor of such an assumption. Pay
attention that the temperature of the phase transition to the incommensurate
state differs strongly for the powder and single-crystal samples ($\sim 19$~K
and $\sim 13.5$~K, respectively). It is known that the reduction of the
crystal, when obtaining the powder sample, brings additional mechanic strains.
Those strains can significantly change the temperatures of phase transitions.
Besides, the transition itself can be ``smeared out'' for some temperature
range \cite{14,15}. Multiferoics are especially sensitive to such external
effects; there phase transitions are related to all their subsystems:
magnetic, electric, and elastic ones.

We performed measurements in single crystal samples. However, in our
experiments the samples cannot be considered as totally ``free''. They were
positioned between two piezoelectric transducers (made of litium niobate), the 
acoustic contact with the latter was provided with the help of a thin ($\sim$ 
1-2~mkm) layer of an organic silicone polymer oil GKZh-94. The feezing of the 
oil at 120~K may introduce to the sample additional strains, which can, in 
principle, change the conditions of the development of phase transitions. 
Besides, strains, applied
to the surfaces of the sample, can be distributed inhomogeneously in its
volume (especially in the case of the large sample 1).

The transition to the magetically ordered phase at $T_N \approx 30.6$~K in 
neodymium ferroborate, as well as the transition to the incommensurate 
structure, is, perhaps, sensitive to the external effects (strains). It can 
be one of the reasons for the existance of such a large difference between the 
presented in literature values for the N\"eel temperature (29-33 K), obtained 
in various
experiments. It is necessary to take into account also the onset below $T_N$
of the spontaneous electric polarization, which, in turn, is also very
sensitive to such effects. It is known that a mechanical strain, applied to
ferroelectrics, totally removed the phase transition \cite{16}.

While the transition to the magnetically orderd phase for all acoustic modes
manifested itself as similar peculiarities at the same temperature, the
transition to the inhomogeneous structure was ``smeared out'' at some
temperature range. It revealed itself in the anomalous change of acoustic
characteristics, which starts below 20~K and finishes about 10~K. The presence
of the temperature hysteresis, characteristic for the transitions to
incommensurate phase, implies that it is of the first order.

Now, let us discuss the features of the behavior of the transverse C$_{44}$
mode in the interval 10~K $\le T \le 30$~K, Fig. 4. The acoustic
``inhomogeneity'' of the sample, which yields to the multi-phase signal of the
sound, is caused, most probably, by the process of the creation of domains in
the antiferromagnetic phase, which agree with \cite{5,9}. We take into
account also that neodymium ferroborate is, probably, the nonintrinsic
ferroelectric \cite{2}, i.e. its electric polarization below $T_N$ is, perhaps,
the consequence of the magnetic ordering, see Fig 4a. In that case
the process of the formation of the domain structure can be more complicated.
For example, the recent paper \cite{17} reported about the observation in
MnWO$_4$ of ferroelectric domains with the unusual property: the electric
polarization was caused by magnetic spirals there. In the sample, where the
domains of magnetic spirals were detected, the variety of antiferromagnetic
domains was observed. It is interesting also that the domain structure was not
periodic, but consisted of macroscopic regions, which differed from each other
by the directions of magnetic and electric vectors. We cannot exclude that in
our sample similar structures appeared, which could be the reason for the
observed multi-phase sound signal.

In the general case, the type of the magnetic (electric) domain structure is
essentially affected by the type of the anisotropy, the orientation of the
crystal surfaces with respect to crystallographic axes, the shape and size of
the crystal, and possible defects. Besides, external effects, like the change
of temperature, elastic strains, etc. also affect on domains \cite{18,15}.

This is why, by changing the shape and the size of a sample, and the glue area
(i.e. unavoidable strains), we affected on the conditions of the formation of
domains. The domain structure, which has been formed in the sample 2,
probably appeared to be more homogeneous than the one in the sample 1. The
conditions of the sound spreading have become more favorable, which yielded
the single-phase signal of the sound, and permitted to obtain smooth
temperature dependences of the sound velocity and attenuation.

\section{Magnetic field dependences of the sound velocity and
attenuation}

Magnetic field dependences of the velocity and attenuation of longitudinal and
transverse modes for fixed values of the temperature in the range 1.7~K
$\le T \le 32$~K have been measured in the interval from zero to 5.5~T. The
field was applied in two directions in the basic plane: along the C$_2$ axis,
${\bf H} \parallel {\bf a}$, and perpendicular to it, ${\bf H} \parallel
{\bf b}$. The field, applied along C$_3$ (${\bf H} \parallel {\bf c}$) has not
yielded any observable changes in the elastic characteristics of the crystal.

Notice from the very beginning, that the most sharp features in the magnetic
field dependences were demonstrated by the transverse modes. The effects in
longitudinal modes were at least one order of magnitude smaller, thus we
analyze only the behavior of transverse modes below. By the way, similar
difference in the behavior of transverse and longitudinal modes we observed in
terbium and praseodymium ferroborates \cite{19,20}, which is, probably, the
generic feature of rare-earth ferroborates.

Let us consider the case ${\bf H} \parallel {\bf a}$. At $T=1.7$~K all studied
transverse velocities show the jump at the field $H_1 \sim$ 0.8~T. The scale of
all observed anomalies was -0.3-1\%. The anomaly in the sound velocity is
accompanied by the step-like increase of the attenuation. The peculiarities in
$H_1$ have the hysteresis character. The example of the field dependence of the
sound velocity and attenuation for the C$_{44}$ mode at 1.7~K is presented in
Fig. 5. The value of $H_1$ is detemined as the average value between the
positions in anomalies for the velocity (attenuation) when increasing and
decreasing the value of the magnetic field. Further growth of the field yields
the increase of the sound velocity (the decrease of the attenuation), which,
perhaps, continues above the maximal in our experiments value 5.5~T.
Such a behavior of acoustic characteristics above $H_1$ is also anomalous.
It is, probably, the consequence of some continuous process, which takes place
in the crystal. We have denoted by $H_2$ the value of the field, at which
the sharp increase of the velocity after the jump in $H_1$ starts, see Fig. 5.

The increase of the temperature reveals some differences in the behavior of
transverse modes. All modes can be divided into two groups, according to the
type of their behavior. The first group is formed by the modes, which develop
in the basic plane with the polarization along ${\bf c}$, and C$_{44}$ mode
(${\bf q} \parallel {\bf c}$, ${\bf u} \parallel {\bf a},{\bf b}$). The second
group is made up of the modes, which wave and polarization vectors lie in the
basis plane (the mode C$_{66}$ and the mode ${\bf q} \parallel {\bf b}$,
${\bf u} \parallel {\bf a}$).

Let us consider the characteristic features of the temperature evolution of
the magnetic field behavior of the groups of both kinds, using as the example
C$_{44}$ and C$_{66}$ modes.The dependences for different temperatures are 
shifted for clarity along the ordinate relative to one another in Figs.6-8.

Mode C$_{44}$ (Fig. 6):
\begin{enumerate}
\item The feature at $H_1$ is reliably registered in the range
1.7~K $\le T\le 20$~K, at that above 13~K it becomes hysteresis-free (line 1).
\item The value of the jump of the velocity at $H_1$ increases with the growth
of the temperature in the range 1.7~K $\le T\le 13$~K. In the range
13~K $\le T\le 20$~K the scale of the feature at $H_1$ becomes smaller. The
slope of the dependences above $H_2$ becomes larger with the growth of the
temperature in the range 1.7~K $\le T\le 20$~K. The most intensive growth is
observed in the range 13~K $\le T\le 20$~K. Above 20~K the slope becomes
smaller.
\item At temperatures $\ge 16$~K in weak fields ($\sim 0.25$~T) one relatively
weak jump appears, which has the hysteresis character, and which can be seen
up to 30~K (line 1').
\item  The feature at $H_2$, almost of the same shape, is registered up to
30~K. With the increase of temperature from 1.7~K till 13~K the feature is
rapidly shifted to weaker fields (line 2), Above 13~K and up to 30~K its'
position is stabilized about 0.6~T (line 2').
\end{enumerate}
Mode C$_{66}$ (Fig. 7)
\begin{enumerate}
\item  The feature at $H_1$ is registered in the range 1.7~K $\le T\le 13$~K
(line 1). For the mode C$_{66}$ the values $H_1$ for each temperature is about
0.1~T below than for C$_{44}$ mode.
\item  The value of the jump of the velocity at $H_1$ and the slopes of
the curves above $H_2$ decrease with the growth of temperature in the range
1.7~K $\le T\le 13$~K. About 13~K the features at $H_1$ and $H_2$ become less
distinguishable. The velocity (and the attenuation) of the C$_{66}$ mode
basically do not depend on the field. Further dynamics of those anomalies in
the range 13~K $\le T\le 30$~K is described in 4.
\item The weak hysteresis anomaly in the velocity in fields $\sim 0.25$~T can
be seen 15~K $\le T\le 18$~K, Fig. 7b. In analogy with the C$_{44}$ mode we
have denoted that region as the line 1'.
\item The feature at $H_2$ is registered up to temperatures $\le 13$~K. Above
13~K the increase of the velocity, which we earlier associated with the
feature at $H_2$ is observed for fields $\le 0.3$~T (for C$_{44}$ mode it is
$\sim 0.6$~T). In the range of temperatures 19~K $\le T \le 29$~K the slope of
the dependences is sharply increased. The hysteresis with the loop ``swing''
$\sim 0.7$~T appears. We think that the C$_{66}$ mode, unlike the C$_{44}$ one,
in this temperature range cannot ``resolve'' the features, related to the
lines 1' and 2'. Hence, we plotted these lines at the lower and upper
boundaries of the ranges of fields, in which the sharp increase of the
velocity is observed (which basically agrees with the boundaries of the
hysteresis loop).
\end{enumerate}

Now let us consider the case ${\bf H} \parallel {\bf b}$. The features at
$H_1$ and $H_2$ can be reliably registered in the temperature range 1.7~K $\le
T \le 13~$K for all transverse modes (lines 1 and 2 of Fig. 8). Note that in
this range for all modes the values of $H_1$ and $H_2$ for each temperature are
about $\sim 0.15$~T higher than in the case ${\bf H} \parallel {\bf a}$. For
the C$_{44}$ mode those features are manifested most sharply, see the inset of
Fig. 8a. The dependences of the C$_{66}$ mode are less prominent. In the range
13~K $\le T \le 20$~K the lines 1 and 2 come together and at temperatures
above 20~K we cannot divide them. In fact, only one line survives, denoted
as 1' (Fig. 8a,b), which we managed to register up to 30~K. The position of the
line 1', basically does not depend on the temperature and it is situated near
$\sim 0.3$~T.

Based on the observed data we have constructed the phase $H-T$ diagrams for
${\bf H} \parallel {\bf a}$ and ${\bf H} \parallel {\bf b}$, Fig. 9. The
points on diagrams are related to the positions of features in the velocity
and attenuation of all studied transverse modes. The lines are guides to eye.

For the analysis of phase diagrams let us use the data of the measurements of
the magnetization \cite{7}, X-ray \cite{5}, and neutron difraction \cite{9,13}.

As we mentioned above, the transition to the incommensurate structure is under
debates. The behavior of the elastic characteristics of NdFe$_3$(BO$_3$)$_4$,
perhaps, permits to clarify the situation to some extend.

Before the neutron data it was believed \cite{7} that below the N\'eel point
the ``easy-plane'' antiferromagnetic collinear magnetic structure exists, which
persist till 2~K minimum. The magnetic moments of iron ions lie in the basis
plane and are oriented along the crystallografic axis ${\bf a}$. Magnetic
moments of neodymium ions lie also in the ${\bf ab}$ plane, magnetized by the
field of iron. In this phase it is possible to form a domain structure, which
permit three types of domains to exist with antiferromagnetic axes, directed
at the angles $2\pi/3$ to each other.

The models of the behavior of the collinear structure in the magnetic field,
applied in the basic plane, are proposed in several studies. They are very
similar to each other. Let us consider one of them, e.g., from \cite{8}. It is
known that when studying the magnetic field behavior of the magnetization at
$T=2$~K weak hysteresis anomalies were observed at the field $H_{sf} \approx
1$~T (for ${\bf H} \parallel {\bf a}$ and ${\bf H} \parallel {\bf b}$),
which were explained as the manifestation of the spin-flop transition \cite{7}.
However, according to \cite{8} that transition takes place according the
following scenario. In the case of ${\bf H} \parallel {\bf a}$ in the field
$H_{sf} \approx 0.7$~T (the value, obtained from the mean-field calculations)
in the domain with the axis of antiferromagnetism along the field direction
the spin-flop transition takes place to the state with magnetic moments, almost
perpendicular to the field. In two other domains with the axes of
antiferromagnetism directed at the angles $\pi/3$ to the field, both magnetic
moments in each domain with different velocity turn to that flop state, see
Fig. 10. The rotation is finished at fields $\approx 1.2$~T.

In the field ${\bf H} \parallel {\bf b}$ the domain with the
antiferromagnetism axis along ${\bf a}$ is already in the flop phase, and its'
magnetic moments turn to the field, Fig. 10. In two other domains the
antiferromagnetism axes are at $\pi/6$ to the direction of the field. Magnetic
moments there with different directions also turn to the state of the
spin-flop phase. At the field $\approx 0.76$~T the system goes to the flop
phase totally, with the jump.

Besides, the calculations have shown that when the temperature increases the
looses of the stability of domains increase, and at 13~K the magnetization
curve looses features, characteristic to the first order phase transition.
Notice that the hysteresis of the magnetization in experimental curves is very
weak even at the lowest temperature 2~K. In the line at 10~K it is almost
undetectable. This is why, it is difficult to determine from the behavior of
the magnetization, how the spin-flop critical field depends on temperature.

On the other hand, the authors of neutron scattering experiments \cite{9}
believe that the collinear antiferromagnetic structure of single crystals
NdFe$_3$(BO$_3$)$_4$ observed in the commensurate phase appears to transform
to an antiferromagnetic long-period spiral below $T_{IC} \approx 13.5$~K
($T_{IC} \approx 19$~K for the powder sample \cite{13}), and the spiral phase
persists at 2~K.

Now, let us return to our results. Note that in both cases
${\bf H} \parallel {\bf a}$ and ${\bf H} \parallel {\bf b}$ one can distinguish
three temperature ranges, at the boundaries of which magneto-elastic
characteristics of NdFe$_3$(BO$_3$)$_4$ change their behaviors: 1.7~K $\le
T < 13$~K (1), 13~K $\le T \le 20$~K (2), and 20~K $\le T \le 30$~K (3).

The magnetic structure of the crystal in zero field in the range (3) is not
under question. At least, the authors of papers, published in 2010 on neutron
scattering \cite{9} and on non-resonant X-ray magnetic scattering \cite{5}
think that it is the collinear ``easy-plane'' antiferromagnet, in which the
domain structure with three types of domains is possible to be formed. Then for
the description of a crystal in a magnetic field, applied in the basis plane
in the range (3) we can use the scenario from \cite{8}. Really, in the case
${\bf H} \parallel {\bf a}$ in the temperature range (3) we have two lines in
the phase diagram: 1' and 2'. The line 1' follows the weak hysteresis jump of
the velocity, that, perhaps, is related to the spin-flop transition in the
domain with the antiferromagnetism axis along the field. The phase between the
lines 1' and 2', according \cite{8} is associated with the rotation of magnetic
domains with the antiferromagnetism axes, directed at $\pi/3$ to the flop
state. Above the line 2', probably, the smooth rotation of all magnetic moments
to the direction of the field (spin-flip) starts, which persists up to fields
$\sim 100$~T \cite{7}. To remind, the line 2' is related to the feature at
$H_2$. It is the value of the field, at which the increase of the velocity
starts, which continues till the maximal in our experiments value 5.5~T.

For ${\bf H} \parallel {\bf b}$ in the temperature range (3) there is only one
line 1' on the phase diagram. We think that it, according to \cite{8}, is
related to the lose of stability of domains with the antiferromagnetism axes,
directed at $\pi/6$ to the field direction, and the transition of the total
system to the flop state. Above the line 1' the smooth rotation of all
magnetic moments to the direction of the field (spin-flip) begins. In this
phase we register the continuous increase of the velocity.

Let us compare the values of critical spin-flop fields $H_1$ for the cases
${\bf H} \parallel {\bf a}$ and ${\bf H} \parallel {\bf b}$ in the range (3):
$H_1({\bf H} \parallel {\bf a}) \approx 0.25$~T, $H_1({\bf H} \parallel
{\bf b}) \approx 0.3$~T. As one can see, the relation
$H_1({\bf H} \parallel {\bf a})/H_1({\bf H} \parallel {\bf b}) =
\cos (\pi/6)$ is fulfilled, which is the evidence in favor of the used model.

Now let us consider the range (1). If there were no transition in a spiral
phase, the magnetic field behavior of elastic characteristics there would be
analogous to the behavior in the range (3), i.e. differences in the behavior
for ${\bf H} \parallel {\bf a}$ and ${\bf H} \parallel {\bf b}$ would exist.
However, the inspection of the phase diagrams show that in (1) diagrams for
${\bf H} \parallel {\bf a}$ and ${\bf H} \parallel {\bf b}$ basically coincide.

Suppose that the spiral phase is realized. As it is shown in \cite{9}, the
spiral is the long-period antiferromagnetic helix, which propagates along the
${\bf c}$ axis with the moments perpendicular to it. The magnetic moments
rotate of about $\pi+\gamma$ around the ${\bf c}$-axis between ajacent
hexagonal planes that are interrelated via trigonal translations (the value of
$\gamma$ increases with the decrease of temperature, and at $T=1.6$~K one has
$\gamma =2\pi/450$). The behavior of such a spiral in the magnetic field has
not been studied in \cite{9}. However, from general grounds one can assume that
in the magnetic field applied in the rotation plane, the spiral-plane flop
transition can be realized (the analog of the spin-flop transition in
collinear antiferromagnets). By the way, similar situation was described in the
recent paper \cite{21}.

The angle of rotation of the magnetic vectors in the spiral is weak, thus the
spiral-flop transition must not be critical to the field direction in the
basis ${\bf ab}$-plane of the crystal. Really, comparing, e.g., the
magnetization dependences on the magnetic field at 2~K for ${\bf H} \parallel
{\bf a}$ and ${\bf H} \parallel {\bf b}$ from \cite{7}, one can see that they
are of the same kind.

The behavior of the elastic characteristics in the interval (3) is also
practically undistinguishable for the cases ${\bf H} \parallel {\bf a}$ and
${\bf H} \parallel {\bf b}$. The most probable reason for it, in our opinion,
is the realization  of the spiral phase in a crystal.

As for the range (2), probably there the co-existence of two phases, the
commensurate (collinear) and incommensurate (spiral) ones takes place. This is
why, we see in phase diagrams in that range simultaneously lines,
characteristic for the low- and higher-temperature phases. By co-existence of
phases one can explain also the hysteresis in the temperature behavior
of the sound velocity and attenuation, which we observed from 10~K till 20~K.

\section{Phenomenological theory}

To describe the observed in our experiments behavior of the magneto-elastic
characteristics of the Nd-based multiferroic we use the Landau-like
phenomenological theory. Let us write the free energy of the magnetic,
electric, and elastic subsystems of the considered system in the form
\begin{equation}
F=F_e + F_m + F_{m-e} + F_{m-el} + F_{el} \ .
\end{equation}
The first term describes the free energy of the electric subsystem,
\begin{equation}
F_e = {a_1\over2} \eta_{\perp}^2 + {a_2\over 4} \eta_{\perp}^4 +
{a_3\over2} \eta_z^2
\end{equation}
where $\eta_z$ and $\eta_{\perp}$ denote components of the electric order
parameter, parallel and perpendicular to the C$_3$ axis of the crystal, and
$a_{1,2,3}$ are the standard coefficients of the Landau series (e.g.,
$a_1 \sim (T-T_e)$, where $T_e$ is the temperature of the electric ordering).
Here we consider the general situation, where the electric polarization can be
caused not only by the magnetic ordering. The second term describes the free
energy of the magnetic subsystem (for simplicity we limit ourselves with the
two-sublattice approximation)
\begin{equation}
F_m = {1\over 2\chi_{\perp}} {\bf M}^2 +{1\over 2\chi_{\parallel}}
({\bf M}{\bf L})^2 -({\bf M}{\bf H}) + F({\bf L}) + F_{anis} \ ,
\end{equation}
where ${\bf M}$ and ${\bf L}$ are the magnetization and the vector of
antiferromagnetism (magnetic order parameter), ${\bf H}$ is the external
magnetic field, $\chi_{\parallel}$ ($\chi_{\perp}$) denote the components of the
magnetic susceptibility of the antiferromagnet, parallel (perpendicular) to
the external magnetic field, $F({\bf L})$ and $F_{anis}$ denote the isotropic
part of the magnetic energy, which depends on the vector of antiferromagnetism
only and the energy of the magnetic anisotropy. The next term describes the
homogeneous and inhomogeneous interaction between the electric and the
magnetic subsystems of the multiferroic
\begin{equation}
F_{m-e} = - D({\bf P} [{\bf M}\times {\bf L}]) +
\alpha_{ijkn}P_iL_j\nabla_kL_n \ ,
\end{equation}
where $\bf P$ is the vector of the electrical polarization, $D$ is the
parameter of the homogeneous magneto-electric interaction, and $\alpha_{ijkn}$
is the tensor of the inhomogeneous magneto-electric coupling. The last two
terms describe the free energy of the elastic subsystem and the
magneto-elastic coupling. By writing this form of the free energy we neglect
the direct interaction between the electric and elastic subsystems. In what
follows we assume the week magneto-elastic coupling, as usually.

It is clear from the Landau expansion of the free energy of the electric part
of the system, the spontaneous electric order parameter has the components
$\eta_z^0=0$ and $\eta_{\perp}^0= \sqrt{-a_1/a_2} \sim \sqrt{(T_e-T)/a_2}$. The
spontaneous electric polarization $P \sim \eta_{\perp}$, according to the
above, has only components, perpendicular to the C$_3$ axis, in accordance
with the measurements of the electric polarization in neodymium ferroborate
\cite{2}.

Then we can denote
\begin{equation}
{\bf H}_{eff} = {\bf H} + D[{\bf L}\times {\bf P}] \ ,
\end{equation}
from which we see that the homogeneous part of the magneto-electric coupling
renormalizes the external magnetic field ${\bf H}$. We know, however, that
magnetic experiments reveal no spontaneous magnetization in the neodymium
ferroborate \cite{2}, hence the vector of antiferromagnetism has to be
parallel to the spontaneous electric polarization at least for weak external
magnetic fields, i.e. the vector of antiferromagnetism has to lie in the
plane ${\bf ab}$, perpendicular to the C$_3$ axis. It agrees with the results
of the neutron scattering in the studied system \cite{9}. It is not difficult
to show using the fact that in the main approximation, neglecting the magnetic
anisotropy, ${\bf M} = \chi_{\perp}({\bf H}_{eff} - x
({\bf H}_{eff}{\bf L}){\bf L})$, where $x=\chi_{\perp}/(\chi_{\perp}
+\chi_{\parallel})$, that for the standard for antiferromagnets situation
$\chi_{\perp} \gg \chi_{\parallel}$ we can approximate the magnetic energy as
\begin{equation}
F_{m} \approx F({\bf L}) + F_{anis} - {\chi_{\perp}\over 2} {\bf
H}_{eff \perp}^2 \ ,
\end{equation}
where ${\bf H}_{eff \perp}$ is the component of the effective magnetic field,
perpendicular to the C$_3$ axis.

We can turn the consideration to the opposite way: let us consider the energy
$F_e + F_{m-e}$ with $a_2=0$. Then, taking into account that
${\bf \eta} \propto {\bf P}$ we can minimize the sum with respect to
${\bf P}$. Such a procedure immediately yields nonzero electric polarization
${\bf P}$, caused by the nonzero magnetic order parameter
$P_{\perp} \sim (D/a_1) [{\bf M}\times {\bf L}] + P_{inh}$, where the last term
is caused by the inhomogeneous electro-magnetic coupling $\alpha$. Even for
${\bf M} =0$ the inhomogeneous coupling between electric and magnetic
subsystems can cause nonzero electric polarization caused by the inhomogeneous
spatial distribution of the vector of antiferromagnetism. It is, probably, the
case for the neodymium ferroborate, which, probably, belongs to the class of
nonintrinsic ferroelectrics \cite{2}.

According to the experiments \cite{9} the inhomogeneous distribution of the
magnetic order parameter is along $z$ (C$_3$) axis. We can write the Landau
series for the magnetic free energy for the components of the order parameter
$L^x = L \cos \phi$ and $L^y = L \sin \phi$ in the weak magnetic field taking
into account the most relevant parts of the magnetic and magneto-electric
energies as
\begin{eqnarray}
&&F_m + F_{m-e} \approx r L^2 + u L^4 + 2w L^2 \cos 2 \phi
\nonumber \\
&&+ V^{-1}L^2 \int d^3x \left[ A (\nabla_z \phi)^2 + \alpha
P_{\perp}\nabla_z \phi \right] \ ,
\end{eqnarray}
where $r = a(T-T_N)$, $T_N$ is the N\'eel temperature, $a$ and $u$ are the
constants of the Landau expansion, $A$ is proportional to the exchange
coupling, $w = K - \chi_{\perp} H^2_{eff\perp}/4$ is renormalized due to the
effective magnetic field coefficient of the magnetic anisotropy ($K$ is the
coefficient of the ``bare'' in-plane anisotropy observed in neodymium
ferroborate \cite{1,2,6,7,8,9}), and the last term in the integrand (the
Lifshitz invariant \cite{22}) is caused by the most relevant part of the
inhomogeneous electro-magnetic coupling of the magnetic order parameter with
the spontaneous electric polarization of the neodymium ferroborate. Here we
have assumed that $L$ is homogeneous, which agrees with the results of neutron
scattering experiments on single crystals \cite{9}. Notice that in neodymium
ferroborate the spontaneous polarization appears only in the magnetically
ordered phase \cite{2}, hence, $T_e =T_N$. Minimizing the free energy with
respect to $L$ and $\phi$ we obtain two equations (cf. \cite{23})
\begin{eqnarray}
&&2Lr + 4uL^3 + AL\left({\partial \phi \over \partial z}\right)^2
\nonumber \\
&&+ 2\alpha P_{\perp} L {\partial \phi \over \partial z} + 4w L
\cos 2\phi = 0 \ , \nonumber \\
&&2AL^2{\partial^2 \phi \over \partial z^2} + 4wL^2 \sin 2\phi =0 \ .
\end{eqnarray}
There are several solutions of this set of equations. First, the solution with
$L=0$ describes the magnetically disordered (paramagnetic) phase. Second,
there is a phase with the homogeneous distribution of $\phi$ with
$\sin 2 \phi =0$ and $L^2 = \sqrt{[a(T_N-T) - 2w]/2u} \approx
\sqrt{a/2u}(T_N-T)^{1/2}$. That solution describes the homogeneous
antiferromagnetic phase. The phase transition between the paramagnetic phase
and the homogeneous antiferromagnetic one is of the second order, and it takes
place at $T=T_N$. Finally, there exists the inhomogeneous for $\phi$ solution.
Neglecting the effective magnetic anisotropy the solution is trivial
$\phi = q_0 z$, where $q_0$ is the vector of a spiral
$q_0 = |\alpha P_{\perp}|/2A$. For $w \neq 0$ we can write the solution as
\begin{equation}
\phi(z) = {\rm am} \left({2\sqrt{w/A}z\over k}, k \right) \ ,
\end{equation}
where ${\rm am}(x,k)$ is the Jacoby elliptic amplitude function with
the parameter $0 \le k \le 1$, which is determined by the parameter of the
effective anisotropy $w$. Substituting the solution to the free energy we can
write
\begin{eqnarray}
&&F_m + F_{m-e} \approx rL^2 +uL^4 -|\alpha P_{\perp}| L^2 {\pi \sqrt{w}\over k
K(k)} \nonumber \\
&&+ L^2 {4w\over 2k^2}\left( k-2+ {4E(k)\over K(k)}\right) \ ,
\end{eqnarray}
where $E(k)$ and $K(k)$ are the complete elliptic integrals of the first and
the second order. Minimizing this expression with respect to $k$ we obtain
\begin{equation}
{k\over E(k)} = \sqrt{wA}{8\over \pi |\alpha P_{\perp}|} \ .
\end{equation}
One can see that the phase of the order parameter can be written
as the Fourier series
\begin{equation}
\phi (z) = qz + \sum_{p=1}^{\infty} {\sin (2pqz) \over \cosh [p\pi
K'(k)/K(k)]} \ ,
\end{equation}
where
\begin{equation}
q= {\pi^2 |\alpha P_{\perp}|\over 8AK(k)E(k)}
\end{equation}
is the vector of the spiral. We see that, as usually, the effective anisotropy
causes the onset of odd harmonics. Obviously, for $w \to 0$ we have $k \to 0$,
hence, $q \to q_0$ and $\phi \to q_0z$. On the other hand, the limit
$k \to 1$ describes the infinitely large period of the spiral structure with
$q \sim \sqrt{w}/2k\sqrt{A}\ln |4/\sqrt{1-k^2}| \to 0$, i.e. it describes
the transition from the inhomogeneous spiral phase to the homogeneous phase
\cite{23} with the jump of the spiral vector, i.e. of the first order. The
free energy of the inhomogeneous phase can be written as
\begin{equation}
F_m + F_{m-e} \approx rL^2 +uL^4 +L^2{\alpha^2 P_{\perp}^2 \over
16AE^2(k)}\left( {4E(k)\sqrt{wA}\over \pi |\alpha P_{\perp}|}
-1\right) \ .
\end{equation}
The magnitude of the magnetic order parameter can be written as
\begin{equation}
L^2 \sim \sqrt{a/2u}(T^* -T)^{1/2} \ ,
\end{equation}
where $T^* \approx T_N - [|\alpha P_{\perp}|/2a](\sqrt{w}/\sqrt{A}
- |\alpha P_{\perp}/2A) \approx T_N - |\alpha P_{\perp}|\sqrt{w}/2a\sqrt{A}$.

In the main approximation the in-plane external magnetic field can produce
the (first order, as a rule) phase transition between the homogeneous and
inhomogeneous phases. By equating the free energies of the homogeneous and
inhomogeneous phases we obtain, e.g., for the weak magnetic anisotropy the
value of the critical field, at which the transition between the inhomogeneous
and homogeneous phases takes place
\begin{equation}
H_c = {1\over \chi_{\perp}}\left(4K - {\alpha^2P_{\perp}^2\over
2}\right) \approx {4K\over \chi_{\perp}} \ .
\end{equation}
On the other hand, from the above we can see that the second order phase
transition between the paramagnetic and the homogeneous antiferromagnetic
phases basically does not depend (in our approximation) on the external
magnetic field, which agrees with our experimental data.

Now, let us turn to the analysis of the behavior of the sound characteristics.
The elastic and the magneto-elastic parts of the free energy can be written as
\begin{equation}
F_{el} + F_{m-el} = {C\over 2}\varepsilon^2 + B\varepsilon L^2\cos 2\phi \ ,
\end{equation}
where $C$ is the elastic modulus, $\varepsilon$ is the distortion, and $B$ is
the magneto-elastic coupling constant. Taking into account that the
magneto-elastic coupling is weak, we can write the equation for the
renormalization of the elastic modulus as
\begin{eqnarray}
&&{\tilde C} = C - {\left({\partial^2 F\over \partial \varepsilon
\partial L}\right)^2\over {\partial^2 F\over \partial L^2}}
\nonumber \\
&&-{\left({\partial^2 F\over \partial \varepsilon
\partial L\phi}\right)^2\over {\partial^2 F\over \partial \phi^2}}
\nonumber \\
&&- {{\partial^2 F\over \partial \varepsilon
\partial L}{\partial^2 F\over \partial \varepsilon
\partial L\phi}\over {\partial^2 F\over \partial L \partial \phi}} \ .
\end{eqnarray}
In the paramagnetic phase we have, obviously, ${\tilde C} = C$. In the
homogeneous antiferromagnetic phase we get
\begin{eqnarray}
&&{\tilde C} = C -{B^2\over 2u} - {B^2L^2\over 2w} \nonumber \\
&&= C -{B^2\over 2u} - {B^2\over 2w} \sqrt{a/2u} (T_N-T)^{1/2} \ .
\end{eqnarray}
Finally, in the inhomogeneous phase we obtain
\begin{equation}
{\tilde C} = C -{B^2\over 2u} \ .
\end{equation}
The sound velocity is proportional to the square root of the elastic modulus.
This is why, the obtained theoretically temperature dependence of the behavior
of the sound velocity is reminiscent to the one, observed in our experiments:
The sound velocity is constant for $T > T_N$, then it decreases with the
decrease of temperature in the homogeneous antiferromagnetic phase, and,
finally, it is almost constant in the inhomogeneous phase for $T < T^*$ with
the value of the velocity lower than the one in the paramagnetic phase.

\section{Conclusions}

We have performed low-temperature ultrasound investigations of the neodymium
ferroborate in the external magnetic field. Based on that investigation we can
make the following conclusions.

In the temperature behavior of elastic characteristics of sound the transition
to the magnetically ordered state is manifested.

The features of the behavior of elastic characteristics of the neodymium
ferroborate in the external magnetic field, applied in the basic plane of the
crystal permit us to suppose that the transition to the incommensurate spiral
phase is realized in the system. This phase transition behaves as the first
order one.

We have constructed the phase $H-T$ diagrams for NdFe$_3$(BO$_3$)$_4$ for the
cases ${\bf H} \parallel {\bf a}$ and ${\bf H \parallel} {\bf b}$.

We have constructed the phenomenological Landau-like theory, which
qualitatively describes the behavior of the sound velocity as the function of
the temperature and the field. The spiral phase is caused by the nonzero
spontaneous electric polarization. The weak in-plane anisotropy modifies the
spiral structure, giving rise to the onset of odd harmonics for the spatial
distrbution of the vector of antiferromagnetism.

Acoustic ``inhomogeneity'' of the sample, which causes the multi-phase behavior
of the signal, and which is revealed in the behavior of the transverse C$_{44}$
mode in the range of temperatures 10~K $\le T \le 30$~K can be the evidence of
the process of the creation of domains in a crystal. To answer the question
whether the domain structure in the form of antiferromagnetic, and, perhaps,
electric domains exists, one needs to perform additional magneto-optic
investigations, analogous to \cite{17}.

\section{Acknowledgments}

The authors thank V.I.~Fomin and V.S.~Kurnosov for the fruitful discussion of
the results of our study. A.A.Z. thanks Max-Planck-Institut f\"ur Physik 
komplexer Systeme, Dresden, for kind hospitality.

\begin{figure}
\begin{center}
\vspace{-0.0cm}
\includegraphics[scale=0.4]{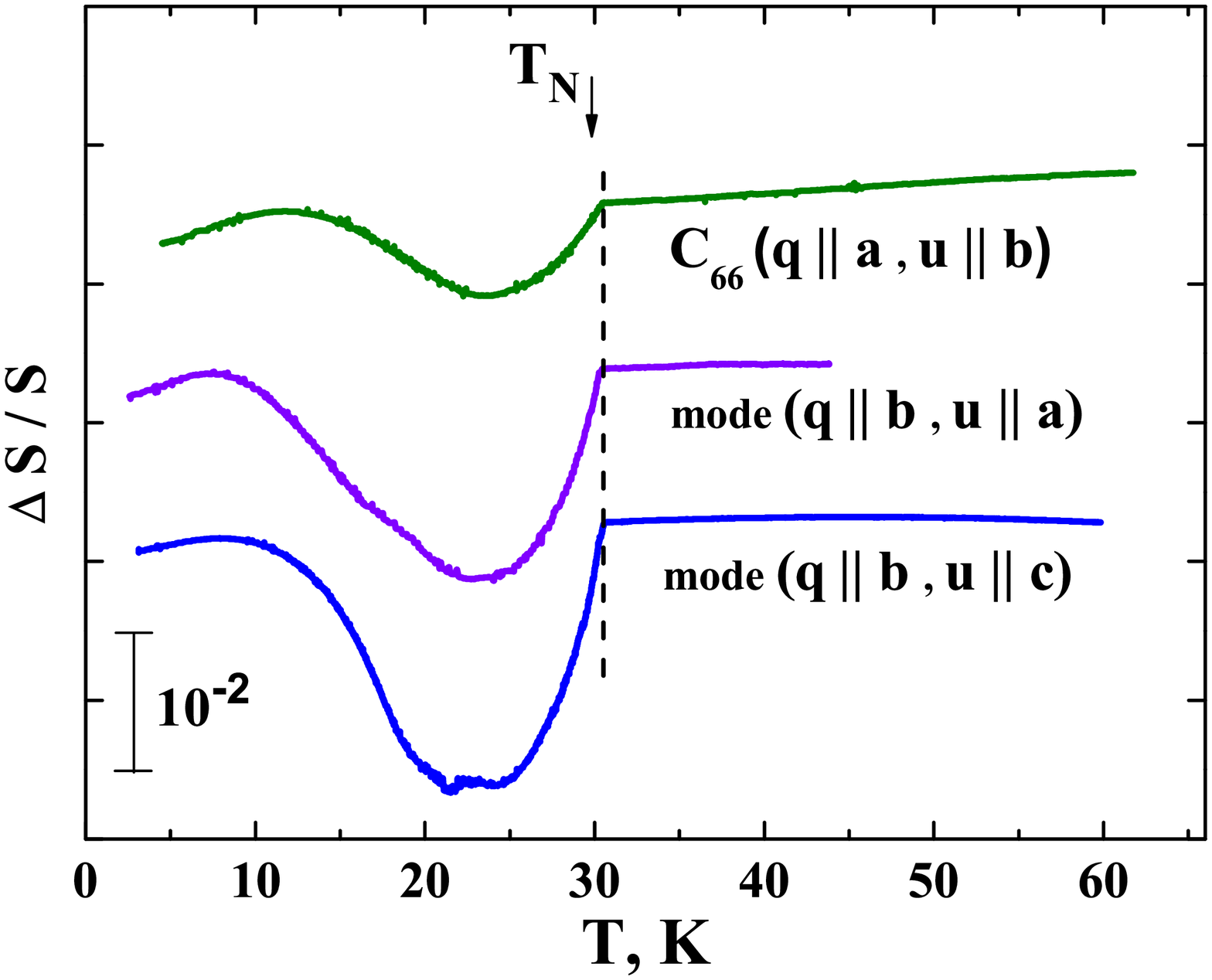}
\includegraphics[scale=0.4]{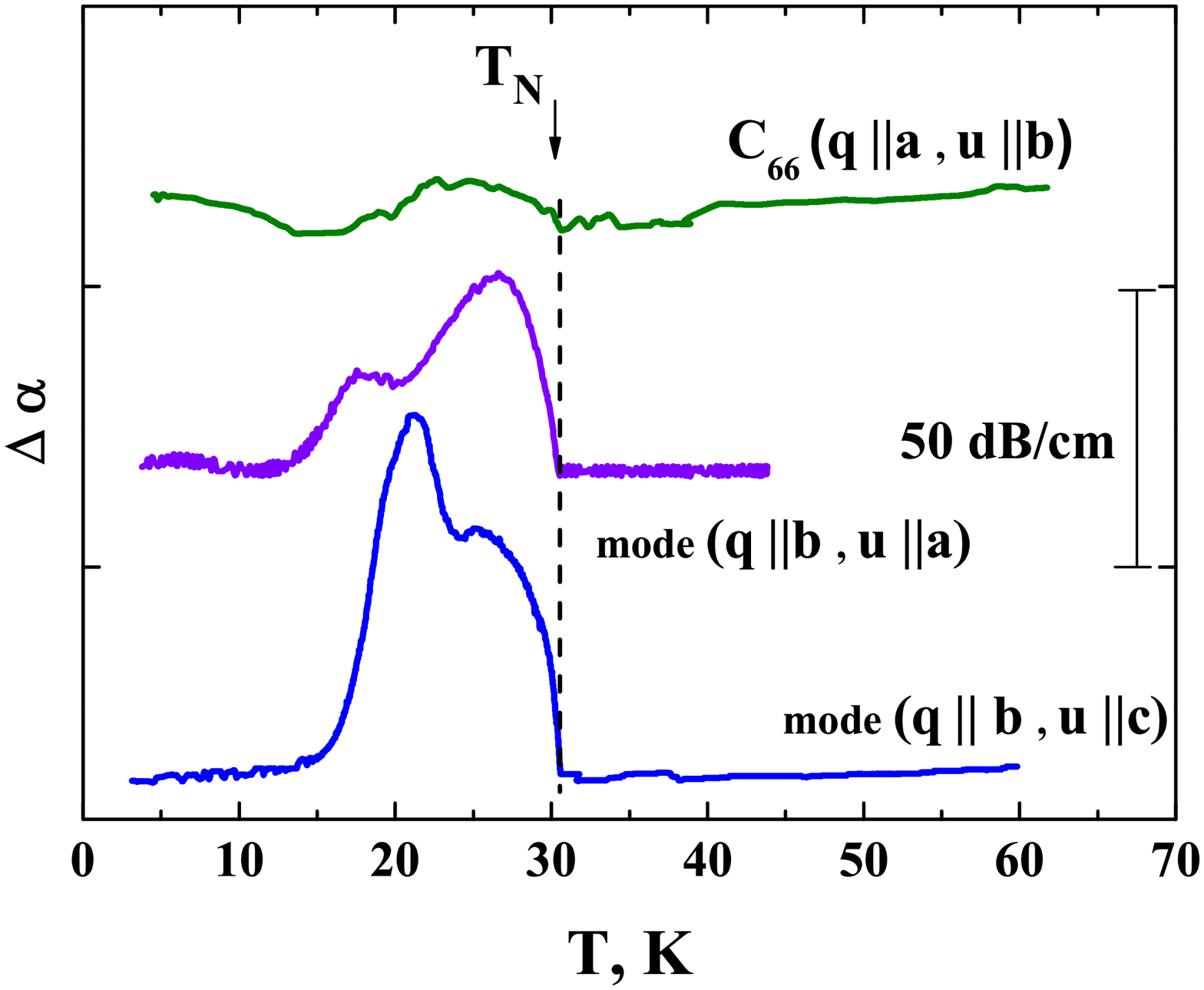}
\end{center}
\vspace{-0.5cm}
\caption{(Color online) The temperature dependence of the behavior of the
sound velocity (a) and attenuation (b) of the various transverse acoustic
modes.}
\label{1}
\end{figure}

\begin{figure}
\begin{center}
\vspace{-0.0cm}
\includegraphics[scale=0.4]{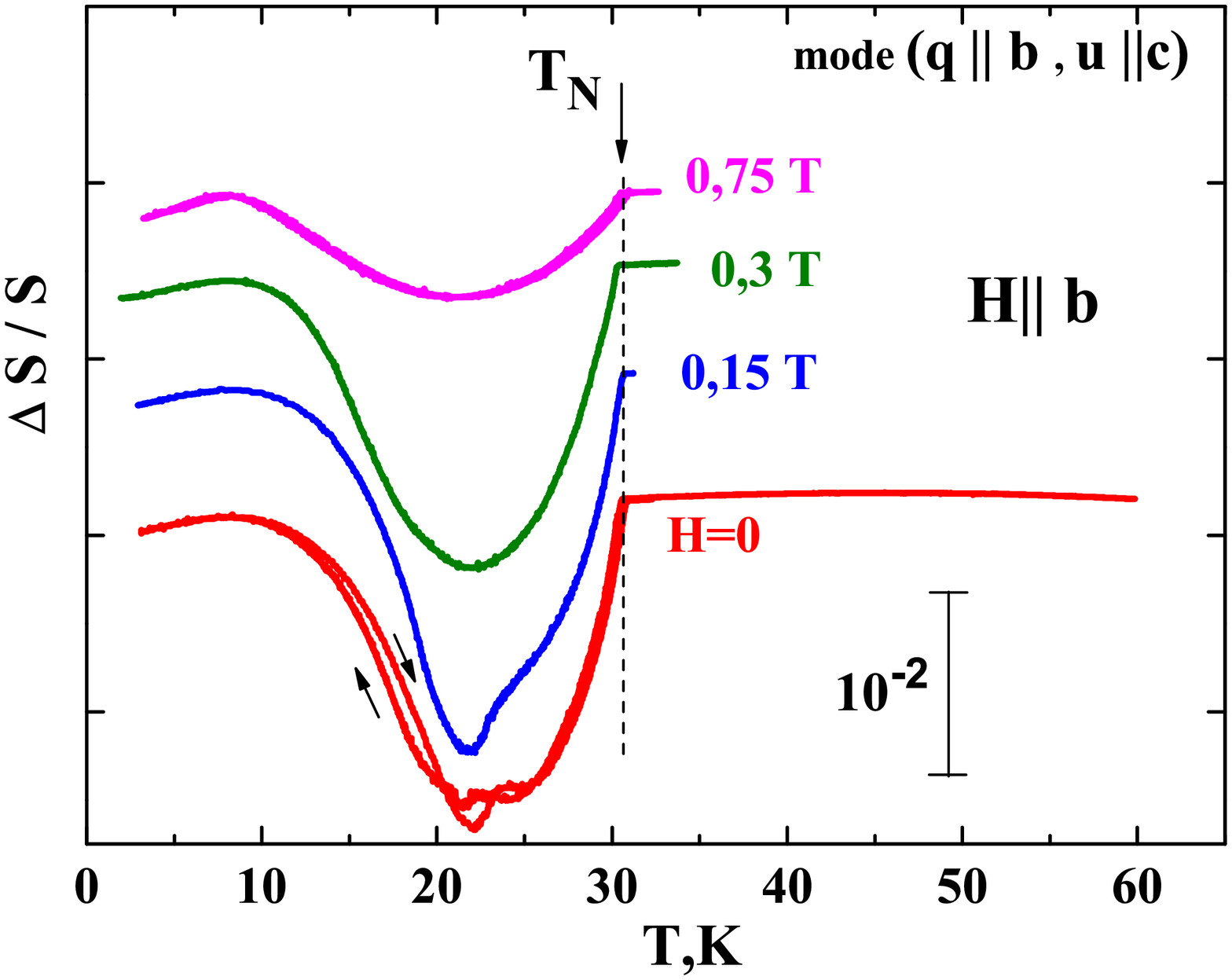}
\includegraphics[scale=0.4]{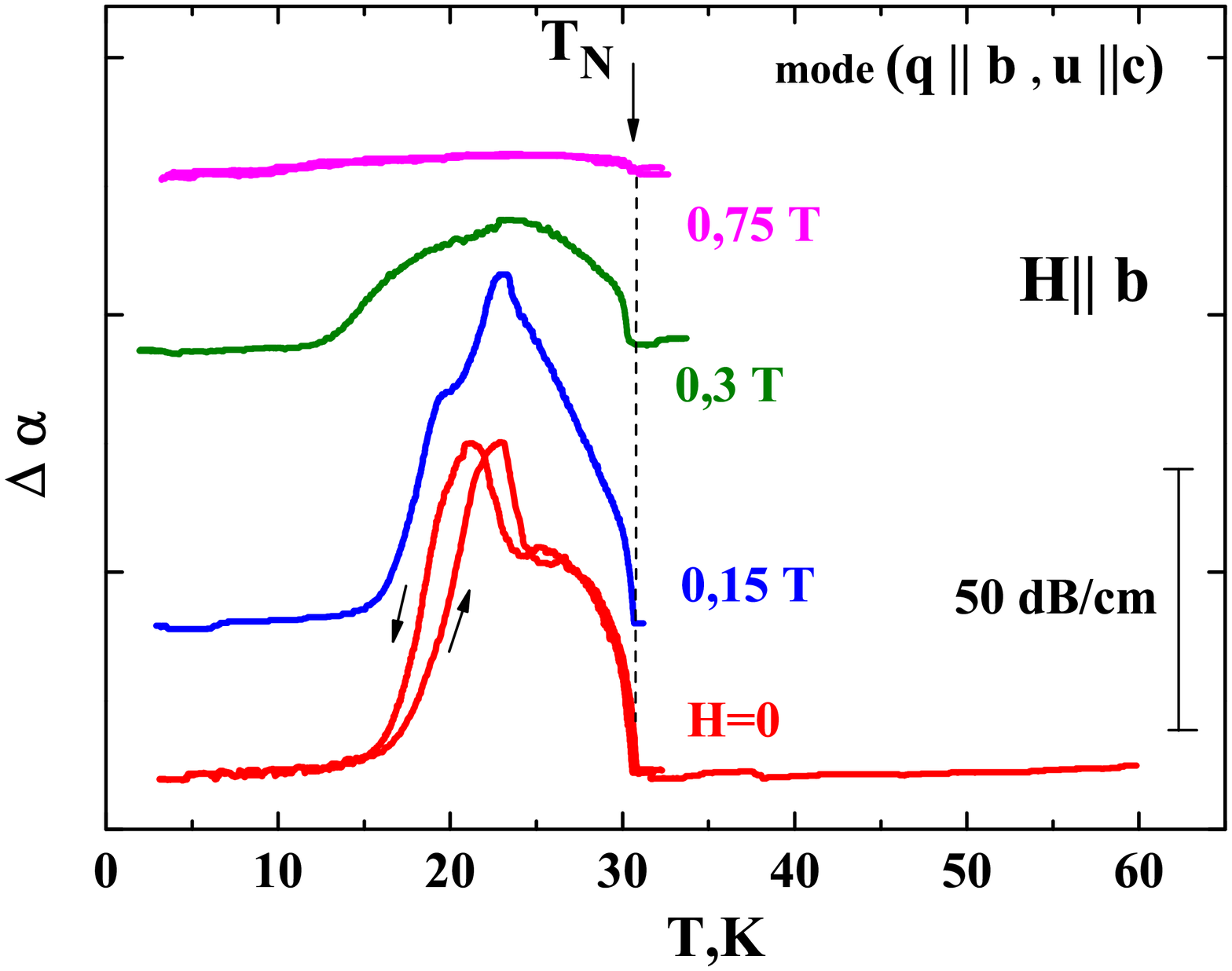}
\end{center}
\vspace{-0.5cm}
\caption{(Color online)The temperature dependence  of the behavior of the
sound velocity (a) and attenuation (b) of the transverse mode (${\bf q}
\parallel {\bf b}$, ${\bf u} \parallel {\bf c}$) in the magnetic field
${\bf H} \parallel {\bf b}$.}
\label{2}
\end{figure}

\begin{figure}
\begin{center}
\vspace{-0.0cm}
\includegraphics[scale=0.4]{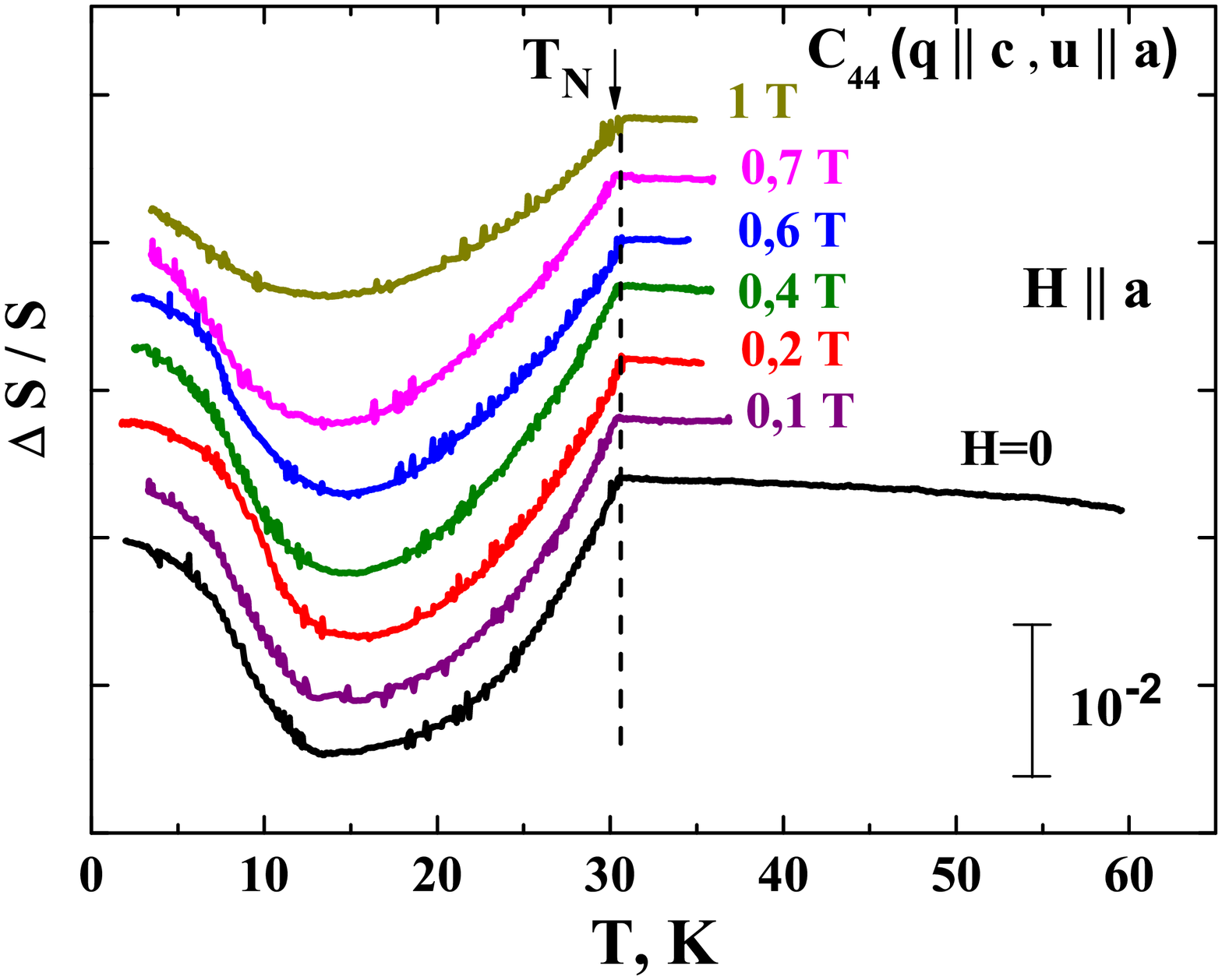}
\includegraphics[scale=0.4]{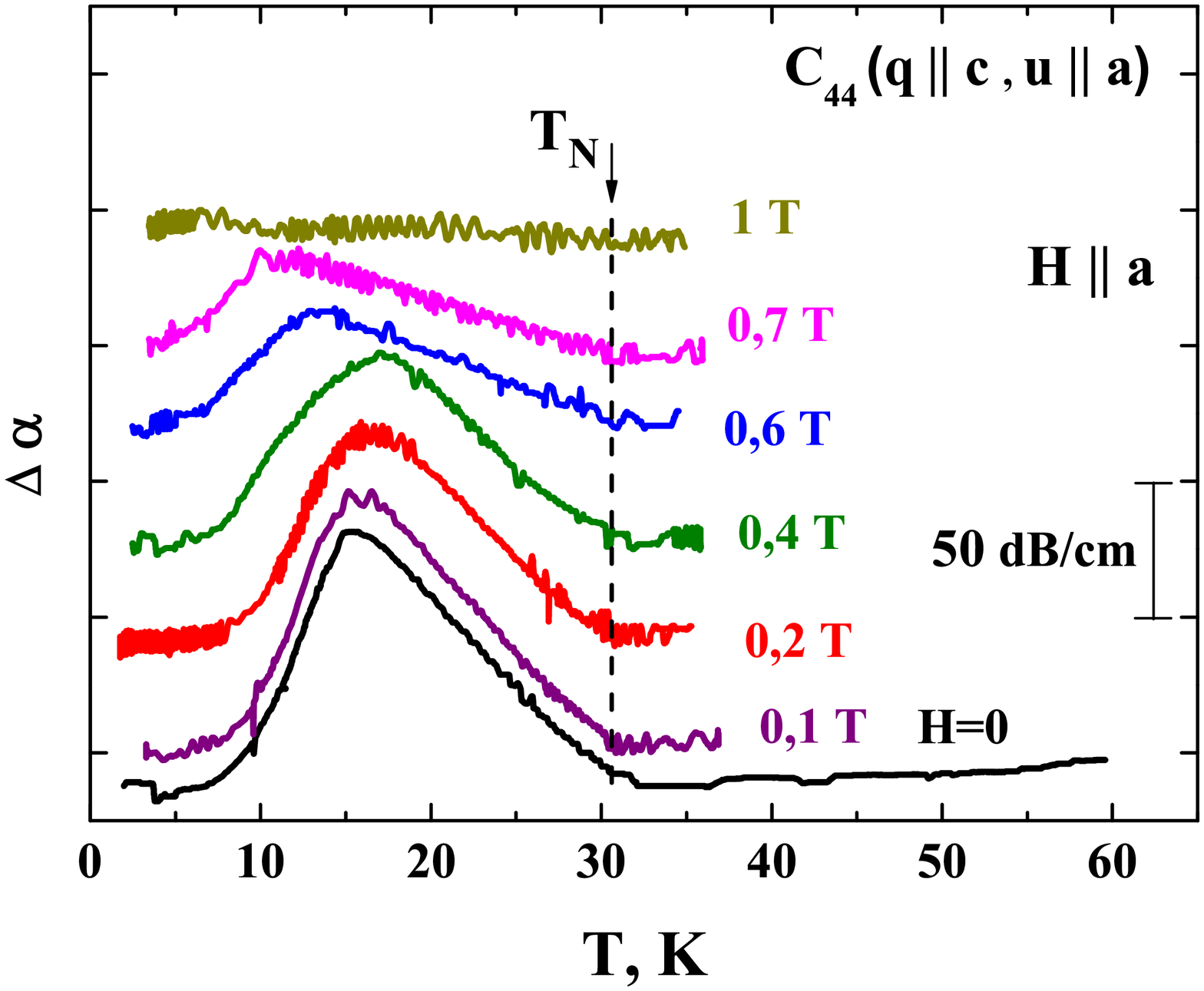}
\end{center}
\vspace{-0.5cm}
\caption{(Color online)The temperature dependence of the behavior of the sound
velocity (a) and attenuation (b) of the transverse mode (${\bf q} \parallel
{\bf c}$, ${\bf u} \parallel {\bf a}$) in the magnetic field ${\bf H} \parallel {\bf a}$.}
\label{3}
\end{figure}

\begin{figure}
\begin{center}
\vspace{-0.0cm}
\includegraphics[scale=0.4]{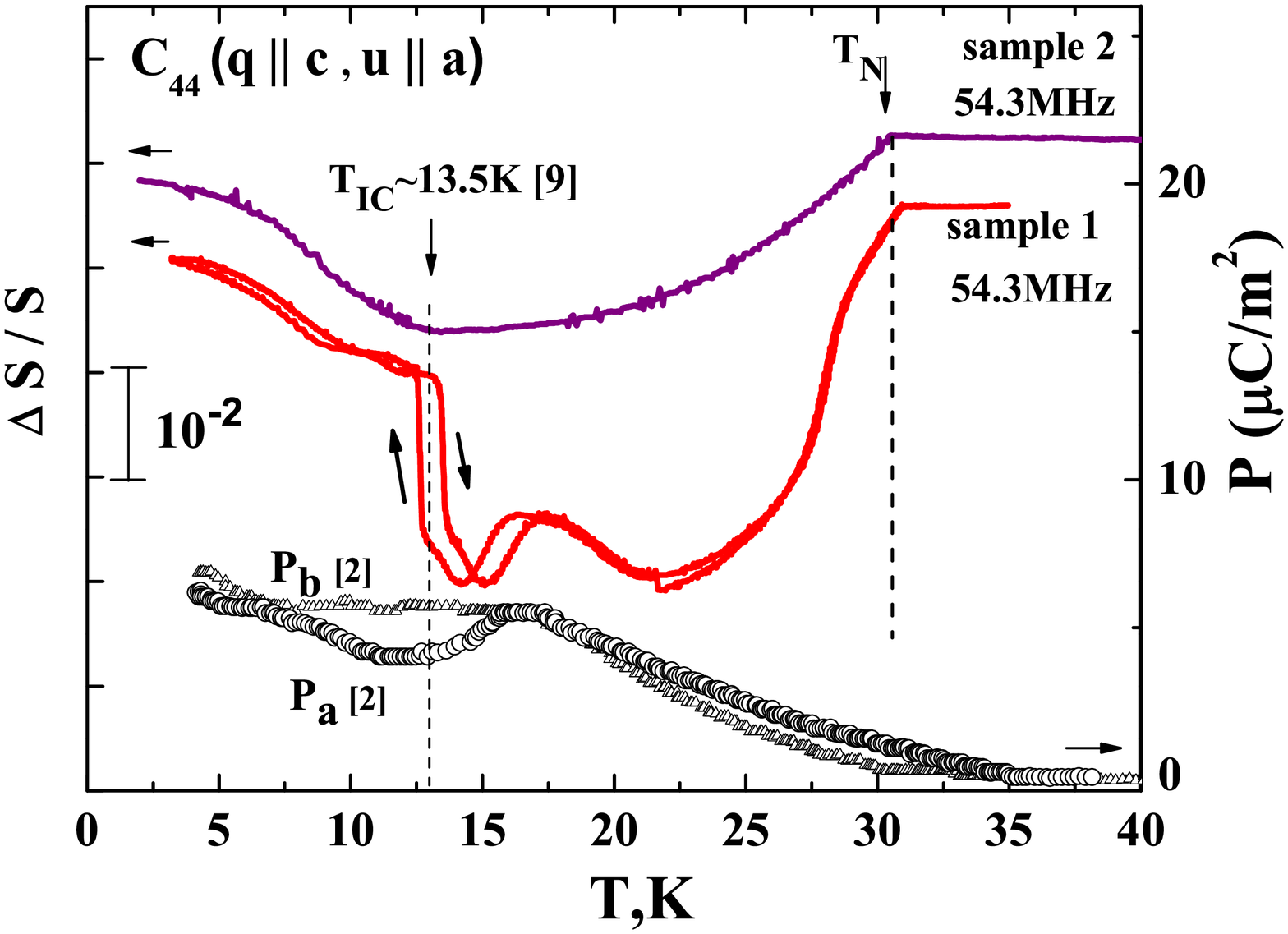}
\includegraphics[scale=0.4]{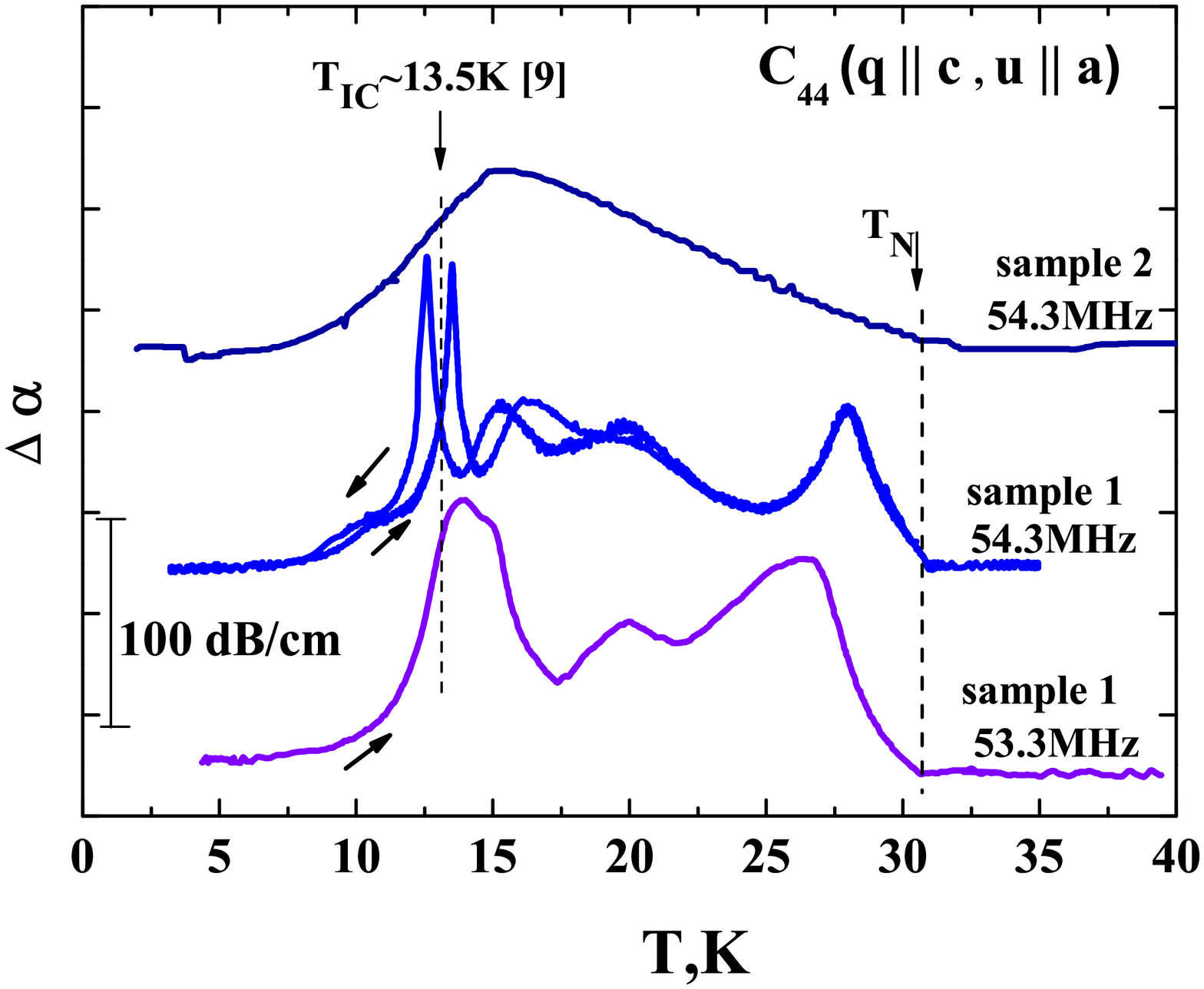}
\end{center}
\vspace{-0.5cm}
\caption{(Color online)The temperature dependence of the behavior of the sound
velocity (a) and attenuation (b) of the transverse mode C$_{44}$ (${\bf q}
\parallel {\bf c}$, ${\bf u} \parallel {\bf a}$) obtained for the samples 1
and 2. The temperature dependences of the spontaneous electric polarization
\cite{2} are also plotted. The temperature dependences of the behavior of the
attenuation for the sample 2 are taken at the frequencies 54.3~MHz and
53.3~MHz.}
\label{4}
\end{figure}

\begin{figure}
\begin{center}
\vspace{-0.0cm}
\includegraphics[scale=0.4]{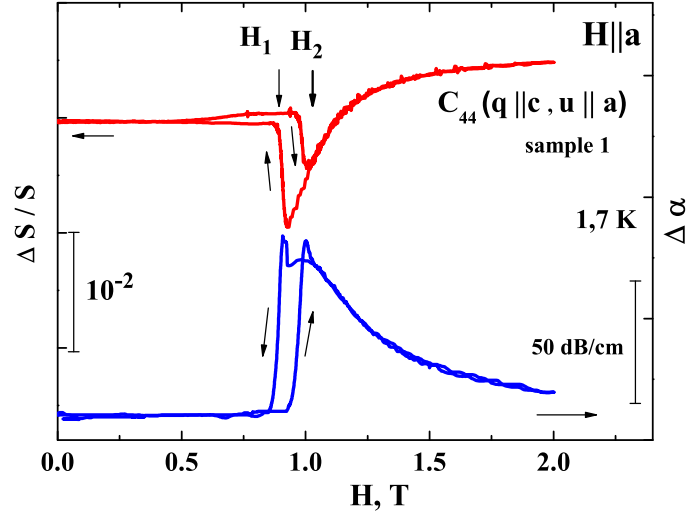}
\end{center}
\vspace{-0.5cm}
\caption{(Color online) Magnetic field dependence of the behavior of the
velocity and attenuation of the acoustic mode C$_{44}$ (${\bf q} \parallel
{\bf c}$, ${\bf u} \parallel {\bf a}$) at the temperature 1.7~K in the
magnetic field ${\bf H} \parallel {\bf a}$.}
\label{5}
\end{figure}

\begin{figure}
\begin{center}
\vspace{-0.0cm}
\includegraphics[scale=1.0]{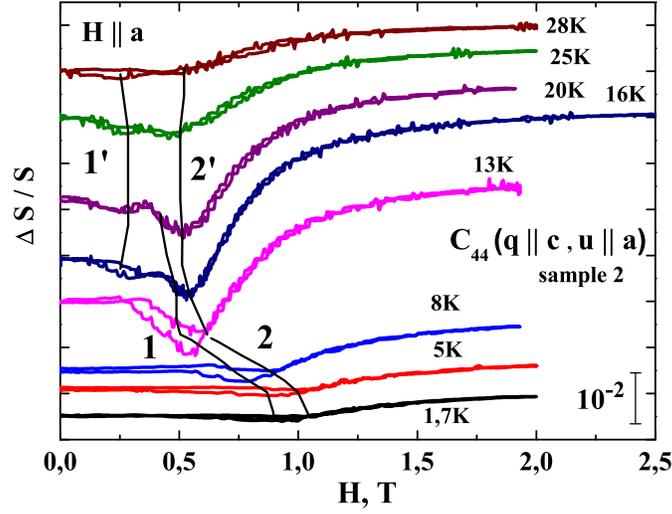}
\includegraphics[scale=1.0]{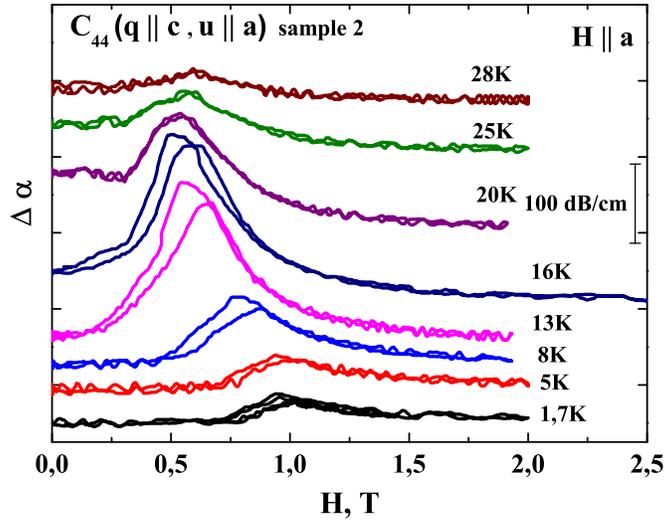}
\end{center}
\vspace{-0.5cm}
\caption{(Color online) Magnetic field dependence of the velocity (a) and
attenuation (b) of the acoustic mode C$_{44}$ (${\bf q} \parallel {\bf c}$,
${\bf u} \parallel {\bf a}$) at various temperatures in the magnetic field
${\bf H} \parallel {\bf a}$.}
\label{6}
\end{figure}

\begin{figure}
\begin{center}
\vspace{-0.0cm}
\includegraphics[scale=0.4]{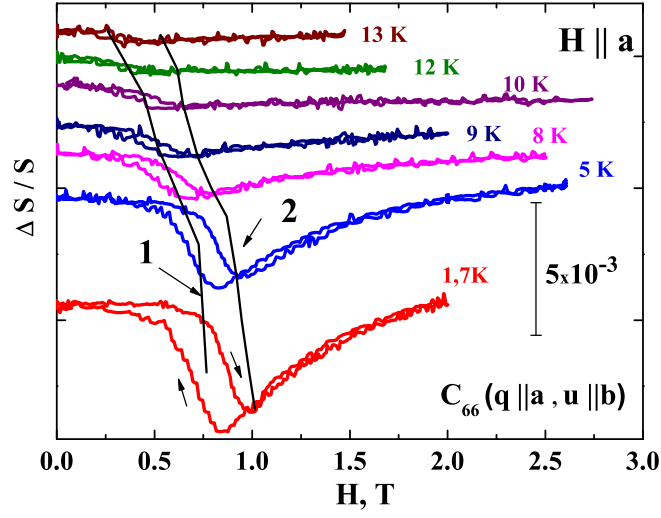}
\includegraphics[scale=0.4]{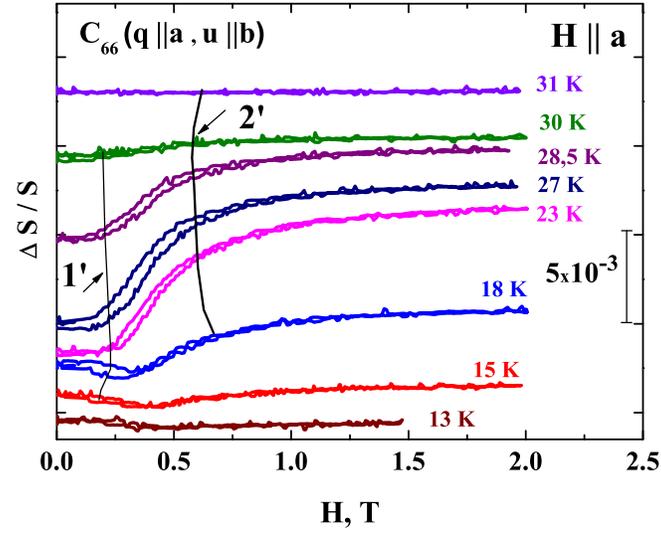}
\end{center}
\vspace{-0.5cm}
\caption{(Color online) Magnetic field dependence of the velocity of the
acoustic mode C$_{66}$ (${\bf q} \parallel {\bf a}$, ${\bf u} \parallel
{\bf b}$) at temperatures from 1.7~K till 13~K (a) and from 13~K till 31~K (b)
in the magnetic field ${\bf H} \parallel {\bf a}$.}
\label{7}
\end{figure}

\begin{figure}
\begin{center}
\vspace{-0.0cm}
\includegraphics[scale=0.4]{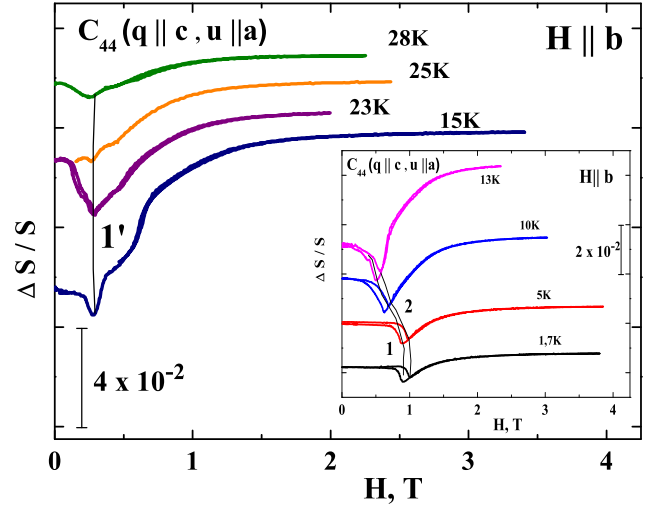}
\includegraphics[scale=0.4]{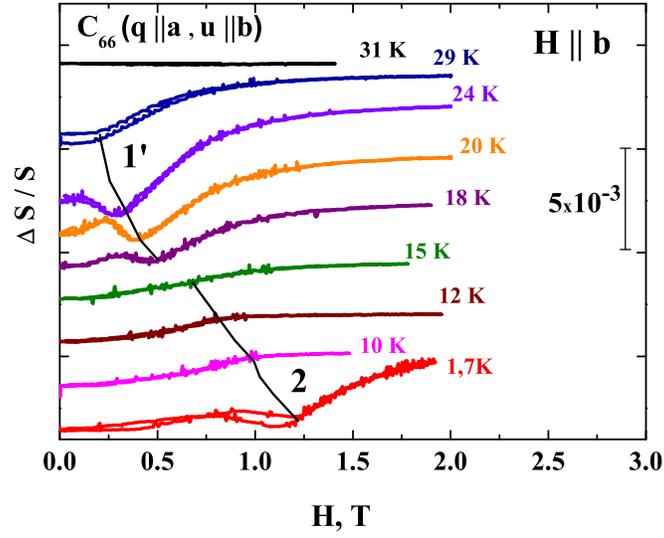}
\end{center}
\vspace{-0.5cm}
\caption{(Color online) Magnetic field dependences of the velocities of the
acoustic modes C$_{44}$ (${\bf q} \parallel {\bf c}$, ${\bf u} \parallel
{\bf a}$) (a) and C$_{66}$ (${\bf q} \parallel {\bf a}$, ${\bf u} \parallel
{\bf b}$) (b) at various temperatures in the magnetic field ${\bf H} \parallel
{\bf b}$.}
\label{8}
\end{figure}

\begin{figure}
\begin{center}
\vspace{-0.0cm}
\includegraphics[scale=0.4]{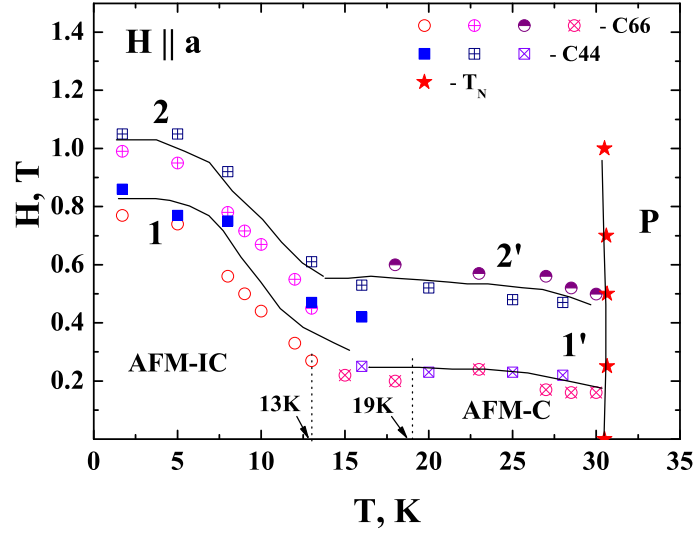}
\includegraphics[scale=0.4]{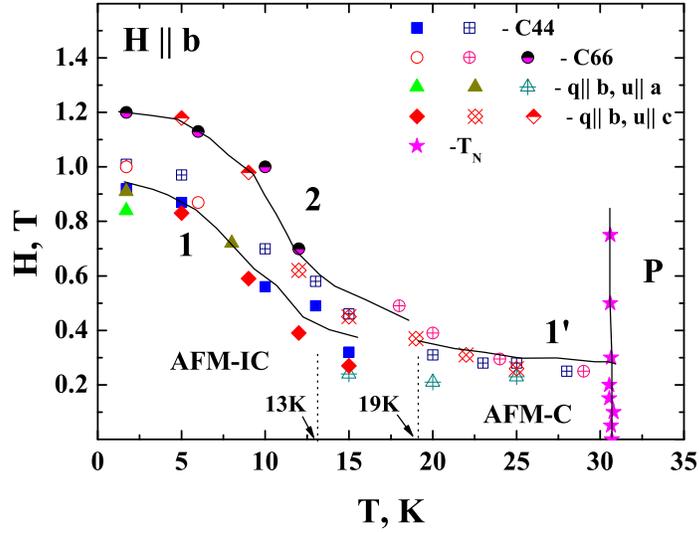}
\end{center}
\vspace{-0.5cm}
\caption{(Color online) Phase $H-T$ diagrams of NdFe$_3$(BO$_3$)$_4$ for
${\bf H} \parallel {\bf a}$ (a), and ${\bf H} \parallel {\bf b}$ (b). The
symbols on diagrams are related to the positions of features in the velocities
and attenuations of transverse acoustic modes. Lines are guides to eye.}
\label{9}
\end{figure}

\begin{figure}
\begin{center}
\vspace{-0.0cm}
\includegraphics[scale=0.4]{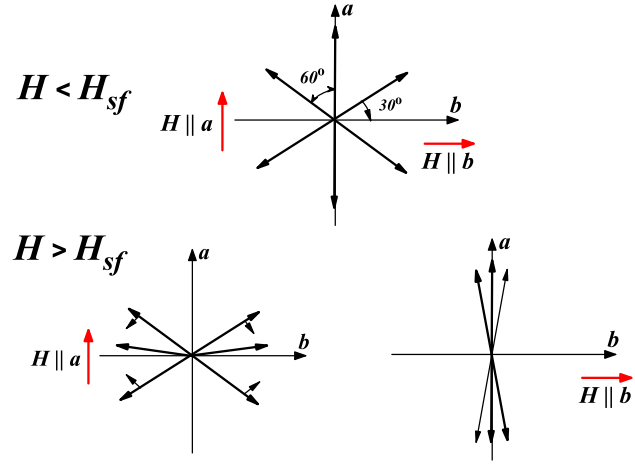}
\end{center}
\vspace{-0.5cm}
\caption{(Color online) The illustration of the model of the spin-flop
transition in the collinear antiferromagnetic phase, the idea is taken from
\cite{8}.}
\label{10}
\end{figure}

\end{document}